\documentclass[10pt]{article}
\usepackage{epsfig}
\usepackage{dcolumn}
\usepackage{bm}
\usepackage{amsmath}
\usepackage{amsfonts}
\usepackage[authoryear]{natbib}
\usepackage{natbib}
\DeclareGraphicsExtensions{.eps,.pdf}

\begin{document}
\begin{titlepage}
\begin{center}
{\bf Modelling Credit Default Swaps:}\\
{\bf Market-Standard Vs Incomplete-Market Models}\footnote{\it{Preprint of an article submitted for consideration in International Journal of 
	Theoretical and Applied Finance © (2014) copyright World Scientific Publishing Company, http://www.worldscientific.com/worldscinet/ijtaf }}\\
	\vspace{12pt}
 {Michael B. Walker}\footnote{Michael B. Walker, Department of Physics, University of Toronto, Toronto, ON M5S 1A7; email: walker@physics.utoronto.ca \label{coords}}$^,$\footnote{The support of the Natural Sciences and Engineering Research Council of Canada is acknowledged.\label{ack}} \\

Version 1: 7 March 2014 \\
Submitted to arXiv: \today	\\
\vspace{2 cm}
\end{center}
\vspace{1 cm}


\begin{abstract} {Recently, incomplete-market techniques have been used to develop a model applicable to credit default swaps (CDSs) with results obtained that are quite different from those obtained using the market-standard model.  This article makes use of the new incomplete-market model to further study CDS hedging and extends the model so that it is capable treating single-name CDS portfolios.  Also, a hedge called the vanilla hedge is described, and with it, analytic results are obtained explaining the striking features of the plot of no-arbitrage bounds versus CDS maturity for illiquid CDSs.  The valuation process that follows from the incomplete-market model is an integrated modelling and risk management procedure, that first uses the model to find the arbitrage-free range of fair prices, and then requires risk management professionals for both the buyer and the seller to find, as a basis for negotiation, prices that both respect the range of fair prices determined by the model, and also benefit their firms.  Finally, in a section on numerical results, the striking behavior of the no-arbitrage bounds as a function of CDS maturity is illustrated, and several examples describe the reduction in risk by the hedging of single-name CDS portfolios.}

\vspace{1cm} \noindent keywords:  credit defaults swaps, CDSs,
hedging, valuation, incomplete markets
\end{abstract}	
\end{titlepage}

\section{Introduction}\label{introduction}
	This article describes the impact that the incompleteness of the market for credit default swaps (CDSs) has on their modelling, hedging, and valuation.   While the general properties of contingent claims in incomplete markets have been extensively discussed in the academic literature (e.g. see the review \citet{sta08}), discussions in both the applied and academic literature that take appropriate account of the incompleteness of the CDS markets are rare, if, indeed, they exist.   A recent article discussing illiquid CDSs \citep{wal12} is an exception. The purpose of the present article is to continue the exploration of issues, begun in the previous article, related to the contributions that incomplete-market methods can make to improving our understanding of illiquid CDSs and their associated risks.  Also, not only  are single CDSs discussed, but the class of problems considered is extended to include single-name CDS portfolios.  
	
Short-comings of the market-standard approach \citep{oka03} are discussed in some detail, and the presentation of the incomplete-market model describes how it changes our view of the properties of credit default swaps. For example, subsection \ref{notMarketable} defines marketability, shows that illiquid CDSs are not marketable, and states that, as a consequence, illiquid CDSs (which are valued in market-standard practice by using the risk-neutral valuation principle) can not in fact be valued correctly by using the risk-neutral valuation principle because they are not marketable.  Other related items can be found in the same subsection, and more details are given in the summary and conclusions, section  \ref{summary}. The background material necessary for treating these subjects is not new, but can be found in the academic literature and in reviews and textbooks treating the subject of incomplete markets (e.g. see \citet{sta08} and \citet{pli97}).  What is new is that the ideas presented below are presented in the context of applying them to a practical model for hedging credit default swaps.  
		
		Numerical examples exploring a number of different properties of illiquid single CDSs and of illiquid single-name CDS portfolios (portfolios that reference only a single name) are an important part of this article.  The first of these numerical examples, discussed in subsubsection \ref{vanillaHedge}, describes the fact that, as a result of market incompleteness, there is no unique price for an illiquid CDS, for the simple reason that there is no replicating portfolio to enforce this price by creating arbitrage opportunities in cases where the illiquid CDS is made available at the `wrong' price.  On the other hand, there is a range of fair prices available to each illiquid maturity. (A fair price is a price that allows neither the buyer nor the seller to acquire an arbitrage profit.)  Subsubsection  \ref{vanillaHedge} introduces a hedge that is called the vanilla hedge.  The vanilla hedge is an approximation to the optimal hedge that was described in detail in \citet{wal12} and is a part of the effort of this article to  make the results of incomplete-market ideas accessible to those with little exposure to incomplete-market methods.  It has the attractive feature that it allows one to determine the no-arbitrage bounds on the prices of illiquid CDSs by elementary methods, and to arrive at analytic formulae for these bounds.  This enables one to obtain an intuitive understanding of the construction of the no-arbitrage bounds and of the behavior of these bounds as a function of the illiquid CDS maturity. A figure showing the richness of detail in the behavior of these no-arbitrage bounds accompanies the explanation.  Since subsubsection \ref{vanillaHedge} is intended to be more or less self-contained, the reader may wish to try jumping ahead to it before continuing.  The selection of a price within the no-arbitrage bounds is, unfortunately, beyond the capabilities of risk-neutral pricing, and involves assessing the risks and the likelihood of profits.  The marking to market of an illiquid CDS is therefore a combined modelling and risk-management problem.

See section \ref{market} for details on the nature of the incompleteness of CDS markets.  Section \ref{SNCDSPort} develops the basic equations used in the incomplete-market approach to the hedging and valuation of single-name CDS portfolios. Most of section \ref{market} and parts of section \ref{SNCDSPort} were described in \citet{wal12}, but some of this material is included here at a faster pace to make the present article more self-contained. Section \ref{numerical} describes several  numerical calculations, and gives figures, that illustrate the use of the results obtained in \ref{SNCDSPort}.  Sections \ref{SNCDSPort} and \ref{numerical} have their own introductions.   

Sources from which the author has profited particularly are the detailed and thorough review of incomplete markets in \citet{sta08}, the elementary introduction to risk-neutral pricing of \citet{pli97}, the presentations of \citet{ros11,ned11}\footnote{\citet{ned11} contains an extensive list of references to recent work on structured credit derivatives, an active area of research closely related to the area of CDSs.\label{ned}}of lessons learned from the 2007-2009 credit crisis, and the appropriateness of a focus on a combined modelling and risk management approach to valuation. Other useful references can be found in \citet{wal12}, as can details that will be of interest to those wishing to develop a numerical implementation, as well as to those wishing to examine results coming from the application of duality theory.

\section{The CDS Market}\label{market}
Standard CDS contracts are currently priced by setting the magnitude of the upfront payment paid by the protection buyer to the protection seller \citep{mar09}.  In addition to paying the upfront price the protection buyer makes quarterly payments of a standardized spread (in the North American market this is 100 bp/yr for investment grade names and 500 bp/yr for high yield names) and in return receives the loss of notional on default of a corporate bond issued by the underlying reference name.

The standard termination dates, as well as the standard premium payment dates, for CDSs are the IMM dates March 20, June 20, September 20, and December 20. Contracts having termination dates at the end of a quarter, say June 20, some integral number of years in the future, will be sold on the market during the quarter leading up to June 20.  During this quarter there will be no liquidity in contracts having one of the other three standard termination dates.  On June 20 there will be a roll when the selling of contracts with a termination date of June 20 will cease, and the selling of contracts having a September 20 termination date will begin.  Furthermore, the number years for which there is liquidity will depend on the name.  For example there may be liquidity only in contracts terminating 1,3,5 7 and 10 years in the future.  Thus, at any given time, there will be liquidity in relatively few of the existing contracts, and it will not be possible to perfectly hedge an arbitrary portfolio of existing CDS contracts of a given reference name.

\section{Single-Name CDS Portfolios in an Incomplete Market}\label{SNCDSPort}
This section develops the basic formulae used in the static hedging of single-name CDS portfolios. Counterparty risk is assumed to have been taken care of by water-tight collateral arrangements. The starting portfolio can be made up of both illiquid and liquid CDSs of arbitrary but given notionals.  At the initial time, the starting portfolio will be hedged using liquid CDSs from the market.  The hedge is a static one and the value of the resulting payoff stream will be followed in time. For simplicity, only the case where the spreads for all maturities are the same (typically the case for post-big-bang CDS contracts) will be treated explicitly in this article. The extension to the case of multiple spreads for each different maturity can be treated by a notational change (see below).   The results developed in this section represent an extension of those in \citet{wal12} from the case of single CDSs to single-name CDS portfolios , but the pace of the presentation is more rapid.  Furthermore,  details of how to approach various technical problems, such as the method of efficiently discretising the continuous paths $(\tau,\rho)$ so as to turn the optimization problem of (\ref{optimization}) into a linear programming problem that can be solved using commercially available software, are given in \citet{wal12}.  
	
Subsection \ref{notation} gives some preliminary definitions and results. Subsection \ref{notMarketable} shows that illiquid CDSs are not marketable.  This result shows that the market-standard use of the risk-neutral valuation principle to value illiquid CDSs has no foundation in mathematical finance theory: the use of the risk-neutral valuation principle to value a contingent claim can be justified only in cases where the claim to be valued is marketable.  Subsection \ref{NAB} develops the main results of this section. These results allow the identification of the range of portfolio prices that are arbitrage-free and those that offer the opportunity for arbitrage profits. They also show how to construct a hedged position that reduces the risk of holding an illiquid CDS portfolio.


\subsection{Preliminary Definitions and Results}\label{notation}
The present time is called $t = 0$.  For this article, the quarterly CDS premiun payment times occurring after $t = 0$, are, beginning with the smallest, $T_1, T_2, ... T_N$, with $N = 21$, so a contract concluded in the quarter preceeding $T_1$ will have a nominal 5-yr maturity at $T_N$. Also, define $T_0 = 0$ (i.e. $T_0$ is time $t = 0$).  The time difference $T_1 - T_0 $ can be less than a quarter of a year, but subsequent time intervals are exactly a quarter of a year. The choice $T_1 - T_0 = (0.25 - \varepsilon)$  yr is made for the numerical examples below.  The quantity $\varepsilon$ is a positive infinitesimal.

Consider a CDS on a given name that defaults at time $\tau$ and has a recovery rate of $\rho$.  All further discussion is for CDSs on the same name.  The quantities $\tau > 0$ and $\rho \in [0,1]$ are random variables and are said to define the path of the CDS.  The present value of the payoff stream of the long-protection unit-notional CDS of termination time $T_m$ that follows path $(\tau,\rho)$ is
\begin{equation}
	\Delta_m(\tau,\rho) = -w {\cal T}_m(\tau) + (1-\rho)d(\tau)U(\tau \le T_m).
		\label{UDef}
\end{equation}
Here
\begin{eqnarray}
	  {\cal T}_m(\tau) & = & \sum_{k = 1}^{I(\tau)-1} (T_k-T_{k-1})d_k
	 	+ (\tau - T_{I(\tau)-1})d(\tau),\ \ 
	 	\tau \in (0,T_m], \nonumber \\
	{\cal T}_m(\tau) &=& {\cal T}_{m,0} 
	\equiv \sum_{k = 1}^{m} (T_k-T_{k-1})d_k, 
	 	\ \ \tau > T_m.
		\label{Udef}
\end{eqnarray}
Also, the function $U(\tau \le T_m)$ is unity for $\tau \in (0,T_m]$ and zero otherwise, the function $I(\tau)$ is defined to be equal to $k$ when $\tau \in (T_{k-1},T_k]$, and the discount factors $d_k$ and $d(\tau)$ are given by $d_k = \exp(-rT_k)$ and $d(\tau) = \exp(-r\tau)$, with $r$ being the risk-free interest rate. In this note, all contracts are considered to have the same spread $w$.  It is also straight-forward to consider the case in which contracts occur that differ from one another in having a number different spreads for each termination date: this is done by including an index $n$ that labels the different spreads that occur for termination date $m$, as in $w_{m,n}$.

It is assumed that there is a liquid market for a small number $K$ of the $N$ different CDS contracts. The p-th liquid contract corresponds to  termination time $T_{m(p)}$,   $p = 1,...,K$.

\subsection{Illiquid CDS Contracts are Not Marketable}\label{notMarketable}
 Consider a CDS contract with a termination time $T_{m0}$ that is not equal to any of the termination times $T_{m(p)}$ of the $K$ contracts forming the liquid CDS market. Such a contract is said to be illiquid. (Note that the use of illiquid here refers to all maturities that are not  explicitly listed as being on the CDS market. This can include CDSs that are identified as marketable or not marketable below.)  A contingent claim X is said to be marketable if a linear combination of the payoff streams of the $K$ contracts on the liquid market can be constructed in such a way that this combination of payoff streams of replicates the payoff stream of the claim $X$. Otherwise the contingent claim is not marketable. The payoff stream of a CDS contract with maturity $T_m$ is given by (\ref{UDef}). It is clear that the illiquid CDS contract with termination time $T_{m0}$ is not marketable.  The discontinuity in $\rho$ when $\tau$ crosses the value $T_{m,0}$ cannot be replicated by the discontinuity in $\rho$ of any CDS contract on the market.  (Note that the mathematical expression for the payoff stream for the illiquid contract with maturity $T_{m,0}$ that is given in (\ref{UDef}) contains the discontinuous function $U(\tau < T_{m,0})$.  It is clear that no linear combination of CDS payoff streams from the liquid market, which have their discontinuities at other maturities, can reproduce the discontinuity from $U(\tau>T_{m,0}))$.  Another way to say this is to say that one cannot form a replicating portfolio for an illiquid CDS contract.  In the same way, it follows that it is not possible to form a replicating portfolio for any single-name CDS portfolio that contains at least one illiquid CDS; such a portfolio is said to be not marketable.

There can be both marketable contracts and contracts that are not marketable in incomplete markets.  If a contract is marketable, there is a replicating portfolio corresponding to this contract. The holder of a marketable contract will be able to short  the replicating contract on the  market in return for the total market price, say $MP$, of the CDSs making up the replicating portfolio.  The contract holder will then be able to combine the marketable contract and the short replicating contract, the combination of the two having no net payoffs, and hence zero value.  The net result of this will leave the marketable contract holder with the cash amount of $MP$, which is the unique value of the marketable contract.

If a contract is not marketable, then by definition there is no replicating portfolio and there is no way to establish a unique, preference-free value for a contract that is not marketable.  However, it is possible to determine an arbitrage-free range of values, any one of which would be considered to be a fair price for the contract. The boundary points of the arbitrage-free range of values are called the no-arbitrage bounds.

The market-standard model for CDSs\footnote{Some knowledge of \citet{oka03}(for the market-standard model) and, say \citet{pli97} (for the valuation of contingent claims in incomplete markets) would be an asset in reading this subsection. \citet{oka03} do not claim to have originated the market-standard model, but rather say that they ``present the market standard pricing model for marking credit default swap positions to market. Our aim is first to explain why credit default swaps require a valuation model, and then to explain the standard model -- the one most widely used in the market.  In the process of setting out the model, we take care to explain and justify the various modeling assumptions made.''  It is known in the academic literature (e.g. see \citet{sta08} that there can be serious inadequacies in market-standard types of arguments, and this section demonstrates some of them in the context of the market-standard model for illiquid CDSs.  In spite of its shortcomings (which the author has not seen discussed in articles by finance practitioners in the credit derivatives area) the market-standard model has continued to be used by practioners (e.g. see \citet{oka08}). More recently, \citet{beu09} examines ``a methodology suggested by Barclays Capital, Goldman Sachs, JPMorgan, Markit (BFJM)/ISDA (2009), (an updated version of this methodology can be found in \citet{isd13}   for conversion of CDS quotes between upfront and running,'' where they cautioned that ``The proposed flat  hazard rate (FHR) conversion method is to be understood as a rule-of-thumb single-contract quoting mechanism, rather than as a modelling device.''  Thus, while there is some appreciation that there are limitations to the usefulness of the market-standard model in its flat hazard rate manifestation, the extent of the difficulties has not yet been discussed in detail.  One of the objectives of this subsection will therefore be to do so.\label{OKAPLI}} has, by construction, a risk-neutral measure $Q$, characterized by having exactly the same number of parameters as there are CDSs (and hence CDS prices) on the market.  Also, the recovery rate is taken to be a given constant.  Calibration to the market prices then uniquely determines the risk-neutral measure $Q$.  Once the unique risk-neutral measure has been determined by calibration, the market-standard practice is to calculate, for an illiquid CDS of maturity $T_{m0}$ and unit notional, a unique mark-to-market price $MtM_Q$ by using
\begin{equation}
		MtM_Q = E_Q(\Delta_{m0}), 
	\label{MtMQ}
\end{equation}
where the right hand side of this equation is the expected value, under the unique market-standard risk-neutral measure $Q$, of the discounted future payoffs for $\Delta_{m0}(\tau)$, given by (\ref{UDef}).   The valuation principle of risk-neutral pricing theory is that, given a risk-neutral probability measure $Q$, and a marketable contingent claim $X$, the time $t = 0$ value of $X$ is given by the expression $E_Q(X)$.  An important condition here is that in order to make correct use of this risk-neutral valuation principle, the contingent claim to be valued in this way must be marketable.  Thus, it is a misapplication of risk-neutral pricing to use (\ref{MtMQ}) to value an illiquid CDS since, as shown above, illiquid CDSs are not marketable. There is thus no sound basis for the interpretation of (\ref{MtMQ}) as a mark-to-market value.  (This article employs the widely accepted practice of using the term `risk-neutral measure'  to describe the probability measure obtained by the procedure of the preceding paragraph, and this practice is adopted here although its suitability is not clear to the author.)

There is another inconsistency in the market-standard model that should be noted.  As described in the previous paragraph, the market-standard model has a unique risk-neutral measure.  However, as stated in \citet{pli97}, ``A model is complete if and only if $\mathbb{M}$ consists of exactly one risk-neutral probability measure.'' Here, $\mathbb{M}$ is the set of all risk-neutral probabilty measures.  Since the risk-neutral measure in the market-standard model is unique, the theorem requires that the model is complete.  However, if the model is complete, one should be able to replicate any contingent claim, and, because it has been shown above that illiquid CDSs are not marketable, one cannot replicate illiquid CDSs.   

\subsection{No-Arbitrage Bounds, Fair Prices, and the Realized Rate of Return for Non-Marketable CDSs and for Single-Name CDS Portfolios}\label{NAB}

This subsection represents an extension of work begun in \citet{wal12}.  In addition to single CDSs, which were considered in the previous article, single-name CDS portfolios are considered,  and further comments on valuation and risk are added.  Consider an existing single-name  portfolio of CDSs, held by a dealer, in which all CDSs are written on the same reference name.  The net notional of all CDSs in this existing portfolio with the same termination time $T_m$ is called $\alpha^{Old}_m$, $m = 1,2,...,N$, where $T_N$ is the largest termination time considered.  The portfolio itself consists of CDSs with $N$ different termination times, and will be called the portfolio $\alpha^{Old}$ (with no subscript $m$).  The portfolio $\alpha^{Old}$ will be assumed to include at least one illiquid CDS, and is therefore not marketable.  (The value of a portfolio containing only CDSs on the liquid market is simply the sum of the market prices of the component CDSs.)

The dealer (also called the seller) can exit the position in $\alpha^{Old}$ by selling this position to a buyer.  The prospective buyer can obtain information about the attractiveness of $\alpha^{Old}$ as an investment by examining the properties of a particular hedged position constructed by adding CDSs from the liquid market to $\alpha^{Old}$. The net notional of the added CDS from the liquid market with termination time $T_m$ is called $\tilde{\alpha}_m$, $m = 1,2,...,N$.  (Readers, please remember that the tilde, as in $\tilde{\alpha}_m$, identifies a CDS on the liquid market.)  The notional $\tilde{\alpha}_m = 0$ by definition if there is no CDS with termination time $T_m$ on the current liquid market. Also, a cash deposit $\beta$ will be added to the dealer's position.  The portfolio of added market CDSs is called $\tilde{\alpha}$, the portfolio $\tilde{\alpha} + \beta$ is called the hedge, and the portfolio $\Delta = \alpha + \beta$, where $\alpha = \alpha^{Old} + \tilde{\alpha}$ is called the hedged position.  It is important to note that the hedged position $\Delta$ will be constructed below using the procedure (\ref{optimization}), and thus satisfies the constraints indicated in that procedure.  The present value of the payoff stream of the hedged position $\Delta$ when the underlying name follows the path $(\tau,\rho)$ is
\begin{eqnarray}
	\Delta(\tau,\rho) & = & \beta + \alpha(\tau,\rho) \nonumber \\
	\alpha(\tau,\rho) &=& \sum_{m=1}^N \alpha_m 
		\Delta_m(\tau,\rho), \nonumber \\
		\alpha_m & = & \alpha_m^{Old} + \tilde{\alpha}_m,\ \ m = 1,2,...,N. 
		\label{Delta}
\end{eqnarray}	
The values of the notionals $\tilde{\alpha}_m$, for maturities that are on the liquid market, as well as $\beta$, are determined by the optimization procedure 
\begin{eqnarray}
	&&V = \min_{\beta,\tilde{\alpha}} \left(\beta +\sum_{m = 1}^Nu_m
	\tilde{\alpha}_m \right) \nonumber \\
	\text{subject to}:\ & &\Delta(\tau,\rho) \geq 0,
	\ \text{for all}\ \tau > 0 \ \text{and}\ \rho \in [0,1].
\label{optimization}
\end{eqnarray}
The quantity $u_m$ in (\ref{optimization}) is defined for CDSs on the market to be the upfront cost of buying unit notional of the CDS with termination time $T_m$. (Recall that, for termination times $T_m$ not on the liquid market, the quantity $\tilde{\alpha}_m = 0$ by definition.)  The quantity $\beta +\sum_{m = 1}^Nu_m\alpha_m$ is the cost of adding the hedge $\beta + \tilde{\alpha}$ to the portfolio $\alpha^{Old}$.  The minimum value of this cost, obtained by the procedure (\ref{optimization}), is $V$. 

Note that there must be at least one payoff stream $(\tau,\rho)$ of the hedged position for which $\Delta(\tau,\rho)$ has the value zero (the minimum value allowed by the constraint of (\ref{optimization})).  This can be seen by supposing that the result of carrying out the procedure (\ref{optimization}) gives a function $\Delta(\tau,\rho)$ that satisfies $\Delta(\tau,\rho) >=c$, where $c > 0$, for all $\tau >0$ and $\rho \in [0,1]$.  Then the quantity $V$ of (\ref{optimization}) can be lowered by lowering $\beta$; thus $c$ must be 0.

Also note that the payoff streams of the hedged position are non-negative for all possible paths $(\tau,\rho)$.  It follows that the expected value $\overline{\Delta}$ of the hedged position (which, for example, in this article will be computed using the physical probability densities described in the appendix \ref{parameters}) must also be non-negative.

The appropriate results for the case where there are no CDSs on the current liquid market are obtained by taking $\tilde{\alpha} = 0$ in (\ref{Delta}) and (\ref{optimization}).

Suppose, now, that a dealer holding the portfolio $\alpha^{Old}$ plans to sell the portfolio to an investor.  The dealer also pays to the investor the cost $V$ of the hedge that the investor can add to $\alpha^{Old}$ in order to form the hedged position $\Delta$.  The investor can thus acquire at zero cost the position $\Delta$ that has only non-negative payoff streams.  At this stage, the investor is able to obtain an arbitrage profit.  However, because of the arbitrage profit, there will be competition for the hedged position under these terms, and the dealer will be able to insist that the investor return some fraction $\lambda,\ \ \lambda > 0,$ and $\lambda$ to be determined, of the expected profit $\overline{\Delta}$.  Expected values here are determined using the physical probability density.  The details of  the simple physical probability density used in numerical examples below are given in the appendix \ref{parameters}. The amount that the dealer can realize from this transaction is thus 
\begin{equation}
	FP(\lambda) = V_{GLB} + \lambda \overline{\Delta},\ \ \ \lambda > 0,
	\label{MtM}
\end{equation}
where $V_{GLB} = -V$ is the greatest lower bound (GLB) on the range of arbitrage-free prices at which the investor can buy $\alpha^{Old}$.  The price $FP(\lambda)$, $\lambda > 0$  is a fair price.  The fact that prices corresponding to $\lambda > 0$ are called fair prices, does not necessarily means that these prices represent good deals for either the buyer or the seller (see below).  The investor can make arbitrage profits by buying at a price lower than or equal to $V_{GLB}$, corresponding to negative $ \lambda$, and acquires a potentially profitable but risky position by buying at a price higher than $V_{GLB}$.     

The portfolio
\begin{equation}
	\Psi = - \lambda\overline{\Delta} + \Delta
	\label{Psi}
\end{equation}
is the investor's net position after acquiring and hedging $\alpha^{Old}$, and paying the amount $\lambda\overline{\Delta}$ to the seller.  Since $\Delta$ is non-negative, with a minimum value of 0, the quantity 
\begin{equation}
		L_{max} = \lambda\overline{\Delta}
		\label{LMax}
\end{equation}
is  the maximum possible loss of the investor's net position (also called the capital at risk).  The maximum possible loss occurs when the path $(\tau,\rho)$ is such that $\Delta(\tau,\rho) = 0$.  A useful interpretation of equation (\ref{Psi})  is that the position $\Psi$ is the result of the investor starting with zero cash and buying the hedged position $\Delta$ at the price $\lambda \overline{\Delta}$. The hedged position $\Delta$, if held until the maximum maturity of the component CDSs, or the default time $\tau$, whichever is sooner, has a realized present value of $\Delta(\tau,\rho)$, and the present value of the realized profit and loss of the investor is $\Psi(\tau,\rho) = -\lambda \overline{\Delta} + \Delta(\tau,\rho)$.  The investor who purchases the position $\Delta$, will have a realized rate of  return  on the capital at risk, $R_T(\tau,\rho)$, often just called the rate of return, which is
\begin{equation}
   R_T(\tau,\rho) = \frac{\Delta(\tau,\rho)-\lambda \overline{\Delta}}{\lambda \overline{\Delta}},\ \ \ \lambda > 0. 
	\label{RT}
\end{equation}
The rate of return is a random variable bearing both default risk (through its dependence on the default time $\tau$) and recovery risk (through its dependence on the recovery rate $\rho$).   The investor's expected rate of return is 
\begin{equation}
	\overline{R_T} =  \lambda^{-1} - 1.
	\label{ERT}
\end{equation}
This equation relates the investor's expected rate of return $\overline{R_T}$ to the price $\lambda \overline{\Delta}$ paid for the position $\Delta$.  Since no-arbitrage constraints require only that $\lambda$ should be positive, these constraints allow a wide range of fair prices $\lambda \overline{\Delta}$.  The dealer who sells the position $\Delta$, and the investor who buys it, must agree on a price $\lambda \overline{\Delta}$ or, equivalently, on the expected rate of return of the package acquired by the investor.

It will be appreciated that it is not just the expected return $\overline{R_T}$ that gives a portfolio its value: the relationship between the expected return and the standard deviation of the return, or other measures of the benefits and risks of the transaction to the participants, should play a role in their determination of what price, within the range of fair prices, would constitut a good deal for them. It is clear that, in order to value derivative positions that are not marketable, and where risk-neutral pricing is not able to define a unique value for the position, disciplines such as risk management may be usefully called upon to contribute to the understanding of value. Furthermore, it is important to note that the absence of a replicating portfolio with which to hedge is an important source of risk.

Given that risk-neutral pricing theory determines only the range of arbitrage-free values allowed for the price of an illiquid CDS, and not a definite price, an investorwill probably want to make use of some of tools developed to measure the risk associated with the hedged position $\Delta(\tau,\rho)$, and hence the risk associated with the rate of return. These tools are illustrated in section \ref{numerical} where the results of a number of numerical calculations are described, including 1) a graphical description of the possible realized values of $\Delta(\tau,\rho)$, 2) a plot of the probability density for $\Delta$ (which also gives the probability density for the profit and loss, $\Psi$), and 3) a figure showing the effect on the riskiness of an illiquid CDS, of reduced liquidity due to a reduction in the number of CDSs on the liquid market .  Section \ref{numerical} has its own introduction to these and other numerical results.

In section \ref{parameters} an expression that is somewhat simplistic is suggested for $P(\tau,\rho)$ for the purpose of carrying out example calculations.  In practice, it will be necessary to undertake a detailed credit analysis of the reference name in question, and to explore a number of scenarios that are the most likely to affect the peformance of  the CDS that is to be valued.  Appropriate probability density functions $P(\tau,\rho)$ should be constructed for each scenario, and results for the different scenarios judged to most likely to affect the performance of the CDS should be taken into account in the final valuation.  For this purpose, the function $P(\tau,\rho)$ could  be extended to include a $\tau$-dependent hazard rate, and a $\tau$-dependent function $\gamma(\rho|\tau)$.  The benefits of integrating scenario analysis and sound risk-management practices into the valuation process were identified in a report \citep{SSG08} issued during the 2007-2009 financial crisis, and based on a survey targeting large firms that had ``concentrated exposure to securitizations of U.S. subprime mortgage-related credit.'' Although the recommendations did not target credit default swaps, CDSs and structured credit products are both credit derivatives, and the recommendations seem to make sense for both markets.  This has provided motivation for a view of the modelling of financial credit derivatives as an integrated modelling and risk management problem, as has been emphasized in \citet{ros11}, for example.

\subsection{Bid and Ask Prices for a single CDS}\label{bidAsk} 
 A dealer who buys a short-protection CDS contract of notional $\alpha^{Old} = -1$ from an investor is said to sell protection to the investor. When the dealer buys this short-protection contract, one of the conditions will be that the investor pay the dealer the amount $V_S$ required to purchase the hedge which, when added to $\alpha^{Old}$, will produce $\Delta_S$, the hedged position associated with unit notional of the short position.  Because the position $\Delta_S$, by itself, will make arbitrage profits for the dealer, an informed investor will insist that the dealer return a fraction $\lambda_S$ of the expected profit, $\overline{\Delta_S}$.  The dealer thus receives the upfront payment 
\begin{equation}
	u_{S}(\lambda_S) = u_{S,0}-\lambda_{S}\overline{\Delta_{S}},\ \ 0 < \lambda_S
\label{uShort}
\end{equation}
As above, this payment is a fair price for $\lambda$ satisfying the given inequality.  Also as above, the participants in the deal will each have to balance considerations of expected return versus risk to make sure that the final  agreed price within the range of fair prices is consistent with their firm's goals.  The quantity $u_{S,0} = V_{S}$ is the least upper bound (LUB) on the arbitrage-free range of ask prices.  Note that the no-arbitrage condition imposed here is $\lambda_S > 0$ and requires that the ask price is always lower than its LUB.  A dealer who buys a long protection contract of $\alpha^{Old} = +1$ from an investor is said to buy protection from the investor.  Arguments similar to those of the previous paragraph lead to the conclusion that a dealer who buys protection from an investor receives the amount $V_L-\lambda\overline{\Delta_L}$ from the investor, or, equivalently, makes the upfront payment
\begin{equation}
	u_{L}(\lambda_L) = u_{L,0}+\lambda_{L}\overline{\Delta_{L}},\ \  0 < \lambda_L
\label{uLong}
\end{equation}
to the investor.  The upfront payment $u_{L}(\lambda_L)$ is called the bid price.  As above, this equation describes a range of fair prices, and further considerations are necessary to fix a final selling price within  the range of fair prices.  Also, $u_{L,0} = -V_{L}$, where $V_{L}$ is the cost of the hedge associated with the long protection contract with notional $\alpha^{Old}= +1$ for the given maturity.  The quantity $u_{L,0}$ is the greatest lower bound (GLB) on the arbitrage-free range of bid prices.  It has been shown \citep{wal12}, by applying the duality theorem, that 
\begin{equation}
	u_{L,0} \leq u_{S,0}.
	\label{duality}
\end{equation}
It was noted above that a dealer who sells protection to an investor receives the payment $u_{S}(\lambda_S)$ from the investor, while a dealer who buys protection from an investor makes the payment $u_{L}(\lambda_L)$ to the investor.  These payments depend on dealer and customer preferences, and one would expect a dealer to negotiate a net profit on the assumption of selling and buying equal notionals.  Thus, one expects the dealer  to set the ask and bid prices such that  $u_{S}(\lambda_S) > u_{L}(\lambda_L)$.  This is a dealer imposed preference argument, and not an arbitrage argument.

Figure \ref{boundsBidAsk} below, shows the results of a numerical calculation of no-arbitrage bounds as a function of CDS maturity. These results show quite striking variation with maturity.  This will be explained in detail in subsections \ref{boundsVsMaturity} and \ref{vanillaHedge}.

\section{Numerical Studies}\label{numerical}
This section groups together figures obtained by numerical calculation using the incomplete-market approach and illustrating various risk- and value-related properties of single-name CDS portfolios.  The figures describe a number of interesting behaviors not found in calculations using the market standard approach. (In fact, the market-standard approach does not even suggest the existence of no-arbitrage bounds, which is the central feature of derivative securities that are not marketable (such as illiquid CDSs). Subsection \ref{singleCDS} describes the calculation of  the no-arbitrage bounds on the $MtM$ values of both liquid and illiquid single CDSs as a function of CDS maturity. Subsection \ref{LMaxReduction} describes the reduction in the risk of holding an illiquid CDS as a result of hedging with CDSs purchased on the liquid market, and gives information on the magnitude of this reduction.  Subsection \ref{PDPL} calculates and shows in a figure the probability density of the profit and loss of a portfolio containing both liquid and illiquid CDSs.  This is an important variable for the characterization of the statistical properties of the hedged position, since it can be used to find the probability of the profit and loss $\Psi$ lying between any two constant values $\Psi_1$ and $\Psi_2$.  Such information will be useful in the assessment of risks. The information one gets, of course, is only as useful as the physical probability density $P(\tau,\rho)$ is accurate.  This subsection also produces a plot  showing $\Delta(\tau,\rho)$ for a single-name CDS portfolio containing CDSs of randomly generated notionals for all quarterly maturities from $1$ to $N = 21$ quarters. The recovery rate in this article is taken to be a random variable, while typically, the recovery rate in the market-standard approach is taken to be a constant known in advance.  Subsection \ref{rhoConstPDPL} calculates the probability density of the profit and loss for the case that the recovery rate is a constant, and finds that the result is qualitatively quite different from that in which the recovery rate is a random variable.  The case where the recovery rate is a random variable better reflects the real situation.  
 
\subsection{CDS No-Arbitrage Bounds and Arbitrage-Free Prices}\label{singleCDS}
\subsubsection{Results from the Optimization Procedure of (\ref{optimization})}\label{boundsVsMaturity}
\begin{figure}   
\includegraphics[scale = 0.7]{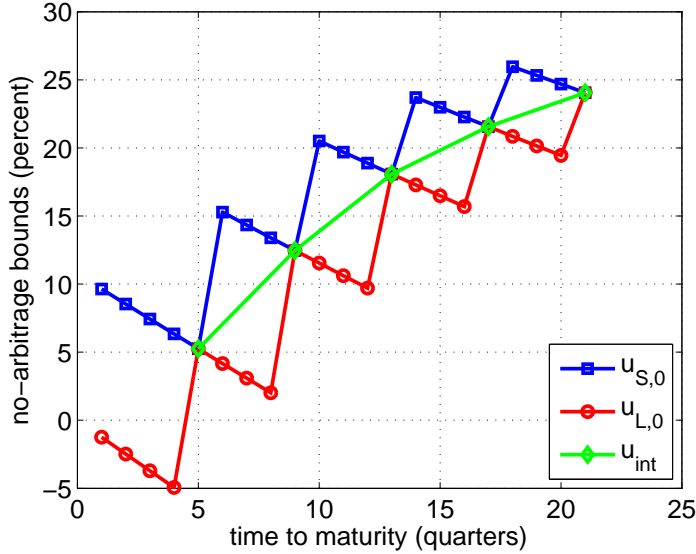} 
\caption{The no-arbitrage bounds for the ask (and bid) prices, labelled $u_{S,0}$ (and $u_{L,0}$), as well as an interpolated upfront price ($u_{Int}$), are plotted as a function of the time to maturity in quarters. The no-arbitrage values of the ask (and bid) prices are less than (and greater than) those of the corresponding no-arbitrage bounds. The equality of the no-arbitrage bounds, and the bid and ask prices holds when the CDS for the given quarter is on the liquid market (i.e. for quarters 5,9,13,17 and 21).  The interpolated line does not extend to maturities less than 5 quarters because there are no liquid CDSs with maturities less than 5 quarters.}
\label{boundsBidAsk}
\end{figure}

A process for determining the arbitrage-free ranges of the bid and ask prices for single CDSs is described in subsection \ref{NAB}.  Some results for the arbitrage-free price ranges for the bid and ask prices of single CDSs described by the parameters in section \ref{parameters} are shown here. The circles labelled $u_{S,0}$ in figure \ref{boundsBidAsk} give the least upper bounds on the ranges of arbitrage-free ask prices ($u_{m,S,\lambda_S}$) for CDSs of maturities $m$ = 1 to 21 quarters. (The index $m$ that denotes the maturity of the CDS is omitted in this discussion, so that $u_{m,S,\lambda_S} \rightarrow u_{S,\lambda_S}$ .) For a given quarter, any price $u_{S,\lambda_S} < u_{S, 0}$ represents a possible arbitrage-free ask price.  Also, any price, $u_{S,\lambda_S} \geq u_{S, 0}$ will yield an arbitrage profit for the dealer who is assumed to be the protection seller.  On the other hand, the squares labelled $u_{L,0}$ give the greatest lower bound on the arbitrage-free range of bid prices and any price $u_{L,\lambda_L} > u_{L, 0}$ represents a possible arbitrage-free bid price.
Recall that  maturities  5, 9, 13, 17, and  21 quarters are on the liquid market (see section \ref{parameters} for the market prices).  For these maturities, the upper and lower no-arbitrage bounds and the market price coincide.  

\subsubsection{A Vanilla Hedge}\label{vanillaHedge}
This subsection explains, using elementary methods, the principal features of the plot of no-arbitrage bounds for an illiquid CDS versus maturity shown in figure \ref{boundsBidAsk}.  The presentation is intended to be accessible to a wide range of individuals with interests in quantitive finance who might be interested in the problem of valuing CDSs.  Because of this, an effort has been made to make  the subsection more or less self-contained, and to reduce the required knowledge of detailed incomplete-market techniques to a minimum.  It has already been shown in subsection \ref{notMarketable} that replicating portfolios cannot be established for illiquid CDSs, and it is the existence of a replicating portfolio that allows arbitrage profits to be made if the price deviates from its unique value, and thus requires the existence of a unique price.  There is thus no unique price for illiquid CDSs.  The best that can be done for illiquid CDSs (or other nonreplicable contingent claims) is to establish a range of fair prices. The basic idea is to establish, for an illiquid CDS called $\alpha^{Old}$, a hedged position called  $\Delta$ that has only payoff streams having non-negative realized present values.  Thus, all of the unhedged risks that would be borne by the buyer of the unhedged illiquid CDS will be eliminated in the hedged position.  The cost of the hedge that, when added to the illiquid CDS to be valued, forms the hedged position, is called $V$.  If this cash amount $V$ is paid by the owner of $\alpha^{Old}$ to the buyer when the contract for $\alpha^{Old}$ is transferred to the buyer, the buyer can use the amount $V$ to buy the hedge. The buyer will then have effectively acquired the hedged position at zero cost.  Because the hedged position has only non-negative payoff streams, the buyer will make an arbitrage profit if there are no further payments. The seller will argue that this is not a fair price, and should be able to find a buyer who will agree to pay back to the seller some positive fraction $\lambda,\ \lambda \geq 0$ of the arbitrage profit $\overline{\Delta}$. The net payment of the buyer to the seller is therefore $V - \lambda\overline{\Delta}$.

This subsection will introduce a  vanilla hedge that  will be able to hedge all of the unhedged risks of the buyer of an illiquid CDS in a very simple manner using only a single CDS from the market.  This will not be the optimum hedge, however, since the optimum hedge will be slightly cheaper and will use all available market CDSs as described in subsection \ref{bidAsk}.  However, the optimum hedge is more complicated, and the procedure described here is simple, easily understood, and, in the examples considered, captures the essential features of the optimal hedge and is close in cost to that of the optimum hedge.  

Now consider this process in more detail.  The owner of the short-protection illiquid CDS contract of unit notional and of maturity $T_M$,  called $\alpha^{Old}_{short}$,  sells it at time $t = 0$ to a buyer.  The buyer takes over this short-protection contract and immediately  (at time t = 0) hedges it  by buying unit notional on the liquid-market, of the long-protection CDS contract (having the same reference name) which is the closest maturity, say $T_{M+}$, in the direction of increasing maturity, to the maturity $T_M$ being hedged. This immediately reduces the unhedged risks of the buyer since now, all risks due to losses on default, as well as risks to variable spread payments, at times between $t = 0$ and $t = T_M$ are perfectly offset by the combination of the two CDSs.  Also, for $t$ between $T_M$ and $T_{M+}$, all losses on default that do occur are associated with the long-protection CDS and represent a profit for the buyer, and not a risk.  There remains the risk that, between these times, the buyer will have to pay the full spread payments associated with the long-protection CDS, which are negative, if there is no default during this time.  The total cost of the hedge for the short-protection illiquid CDS contract of maturity $T_M$ is therefore
\begin{equation}
		V_S(M)= u(T_{M+}) + w \left({\cal{T}}_{M+,0}- {\cal{T}}_{M,0} \right)
\label{hedgeCostS}
\end{equation}
where $u(T_{M+})$ is the upfront cost of the long-protection liquid CDS of maturity $T_{M+}$, the term $ w \left({\cal{T}}_{M+,0}- {\cal{T}}_{M,0} \right)$   is a cash contribution to the hedge to compensate for the potential negative spread payments. Also, ${\cal{T}}_{m,0}$ for $m = M+,M$ is defined in (\ref{UDef}), and the cash contribution proportional to $w$ in (\ref{hedgeCostS}) is approximately linear in $(T_{M+}-T_M).$  The cash payment just referred to is what gives rise to the term $\beta$ in (\ref{Delta}).  The two CDSs and the cash deposit $\beta$ make up the the hedged position $\Delta$ that has only non-negative realized present values for its payoff streams.  Suppose that the illiquid contract to be sold, $\alpha^{Old}_{short}$, is transferred to the buyer, and in addition, the seller gives to the buyer the cost of the hedge, $V_S(M)$.  Then the buyer acquires, at zero cost, a hedged position having only non-negative payoffs.  This would allow the buyer to make an arbitrage profit, and effectively guarantees the sale of the CDS, since there would be expected to be strong demand for securities that give arbitrage profits.  In fact, the demand for such securities should result in a buyer being found that would be willing to return a certain fraction (say $\lambda$) of the expected profits $\overline{\Delta}$ from the hedged position. (The amount returned to the seller is called $\lambda\overline{\Delta}$, following subsection \ref{bidAsk}).

 Now consider an investor (the seller) who sells at time $t=0$ a long-protection illiquid CDS called $\alpha^{Old}_{long}$ of unit notional and of maturity $T_M$. The no-arbitrage bound for this case is again found by constructing a hedged position $\Delta$ that has only non-negative payoffs.  To construct a hedged position with only non-negative payoffs, one begins by buying unit notional on the liquid-market, of the short-protection CDS contract which is the closest maturity, say $T_{M-}$, in the direction of decreasing maturity, to the maturity $T_M$ being hedged. This immediately reduces the unhedged risks of the buyer since now, all risks due to losses on default as well as spread payments at times between $t = 0$ and $t = T_M$ are perfectly offset by the combination of the two CDSs.  Also, for $t$ between $T_{M-}$ and $T_M$ all losses on default that do occur represent a profit for the buyer, and not a risk.  There remains the risk that the buyer will have to pay the full spread payment, which is negative, between times $t = T_{M-}$ and $t = T_{M}$ if there is no default during this time.  The total cost of the hedge for the long-protection illiquid CDS of maturity $T_M$ is 
\begin{equation}
		V_L(M) = - u(T_{M-}) + w \left({\cal{T}}_{M,0} - {\cal{T}}_{M-,0}\right)
\label{hedgeCostL}
\end{equation}
where $-u(T_{M-})$ is the upfront cost of the short-protection liquid CDS of maturity $T_{M-}$, and the term proportional to $w$ is a cash payment from the seller to the buyer to compensate for the potential negative spread payments.  The two CDSs and the cash deposit $\beta$ make up the position $\Delta$ that has only non-negative realized present values for its payoff streams.

Note that for (\ref{hedgeCostL}) to be applicable to the maturity $T_M$, there must be a liquid maturity on the market that is shorter than $T_M$.  This is not the case for maturities $T_M$ less than $M = 5$ quarters in figure \ref{boundsBidAsk}.  For this case, the loss on default of the long-protection CDS that one wants to hedge always represents a non-negative payoff for the holder of the illiquid contract.  Only the spread payments that occur in cases when there is no default need to be offset so that the buyer will take over, at zero risk, a position with only non-negative payoffs. The cost of these offsetting spread payments is
\begin{equation}
	V_L(M)= w{\cal{T}}_{M,0},\ \ M = 1,2,3,\ \text{or 4 quarters}.
	\label{hedgeCostLSmall}
\end{equation}

The notation used to describe the no-arbitrage bounds in subsection \ref{bidAsk} and in figure \ref{boundsBidAsk} is $u_{S,0}(M) = V_S(M)$ and $u_{L,0}(M) = -V_L(M)$, except that the $M$ is suppressed in the figure..  The analytical expressions given for $V_S(M)$ and $V_L(M)$ in equations \ref{hedgeCostS}, \ref{hedgeCostL}, and \ref{hedgeCostLSmall} enable one to understand the main qualitative features of these bounds as plotted in figure \ref{boundsBidAsk}.  The upfront prices for the available liquid CDSs can be read off figure \ref{boundsBidAsk} for the maturities 5, 9, 13, 17, and 21 quarters.  As an example, consider the illiquid CDSs having maturities 14, 15, and 16 quarters, which are bracketed by the liquid maturities 13 and 17 quarters.  The ask-price no-arbitrage bounds for the 14, 15, and 16 quarter maturities lie on an approximately straight line anchored by the market price of the 17 quarter liquid maturity ($T_{M+}$) CDS.  As the illiquid maturity in question (16, 15, and 14) gets further from this anchor, the contribution to ask-price no-arbitrage bound grows linearly with decreasing maturity (e.g. see (\ref{hedgeCostS})).  The magnitude of the increase in the no-arbitrage bound with decrease in illiquid maturity starting from a given liquid maturity anchor point is nearly the same for each of the liquid maturities taken as the anchor point.  (There is a slight magnification of this increase at lower liquid maturity anchor points due to the fact that the effect of the discount factors $\exp(-r_Ft)$ is larger.)  The situation is very similar for the bid price no-arbitrage bounds, except that the anchor points for these are liquid-maturity prices lying to the left of the illiquid maturities of which one wants to find the no-arbitrage bounds.

It is important to note that the no-arbitrage bound do not behave as if they were interpolated between the prices of liquid CDSs, but rather, in the example considered, the illiquid ask price bounds are extrapolated to lower maturities from the nearest higher maturity liquid CDS market price, and the illiquid bid price bounds are extrapolated to higher maturities from the nearest lower maturity liquid CDS market price.

\begin{table}

\centering
\begin{tabular}{|c||c|c|c||c|c|c|c|}
\hline
$T_M$ & 10 & 11 & 12 & 14 & 15 & 16 \\
\hline
$u_{S,0}$ (Opt) & 20.50 &  19.69 & 18.88 &  23.70 &  22.98 &  22.27 \\
\hline
$u_{S,0}$ (Van) & 21.61 & 20.43 & 19.25 & 25.02 & 23.86 & 22.71 \\
\hline
$u_{L,0}$ (Opt) &  11.5425 & 10.6197 & 9.701  & 17.2789 & 16.4817 & 15.6886   \\        
\hline
$u_{L,0}$ (Van) & 11.28 & 10.10 & 8.92 & 16.91 & 15.75 & 14.60 \\
\hline
\end{tabular}
\caption{\label{vanillaBounds}Vanilla (Van) values for No-Arbitrage Bounds Versus Those from Optimization (Opt).}
\end{table} 

Table \ref{vanillaBounds} gives the magnitude of the optimum no-arbitrage bounds for seleted maturities, as well as the magnitudes for the corresponding no-arbitrage bounds calculated from the vanilla hedge.  These are presented in a table rather than a figure because the values are sufficiently close in many cases that there is an overlap of the markers indicating the positions of  the two values.  Note that the cost of the hedge giving the no-arbitrage bound on the ask price ($u_{S,0}$ (Opt)) calculated by the optimization procedure) is always lower than that calculated using the vanilla hedge.  This is expected since the optimization procedure is designed to give the hedge of lowest price $V_S$ constructed using all available liquid CDSs,  and $u_{S,0} = V_S$.  Similarly, the optimum no-arbitrage bound $u_{L,0} = -V_L$ on the bid price obtained by minimizing the cost $V_L$ of the hedge with respect to the notionals of all available market CDSs is greater that of the no-arbitrage bound calculated using the vanilla hedge.  All of this is as expected.  

Figure \ref{boundsBidAsk} shows a line, labelled by diamonds and called the interpolated line, that is a linear interpolation between the known market prices for illiquid CDSs.  The interpolated line ends at a time to maturity of 5 quarters because there is no market price to interpolate to having a time to maturity of less than 5 quarters.  Note that the no-arbitrage bounds on the ask and bid prices are not all placed symmetrically with respect to this line.  The no-arbitrage bounds for the ask-price and bid-price  are equidistant from the interpolated line for the maturity $T_M = 15$ quarters.  However, the no-arbitrage bound for the ask price is approximately 3 times further from the interpolated line than is the bid-price no-arbitrage bound for the maturity $T_M = 14$ quarters, and the bid-price no-arbitrage bound is approximately 3 times further from the interpolated line than is the ask-price no-arbitrage bound for the maturity $T_M = 14$ quarters.  These details could well have significant consequences for the determination of the ask and bid prices themselves.

\subsection{Reduction of the Maximum Possible Loss by Hedging}\label{LMaxReduction}
The maximum possible loss is an important measure of the risk of the position $\Delta$ (or $\Psi$) acquired by a buyer of a single-name CDS portfolio. This subsection investigates the effect of  hedging on the value of $L_{Max}$ for a number of portfolios.  The finding is that the hedging procedure (\ref{optimization}) can substantially reduce the maximum possible loss $L_{Max}$, and, not surprisingly, that the more CDSs that there are on the liquid market, the greater the reduction in $L_{Max}$.  Also, if it is agreed that one of the definitions of increased liquidity is that the number of different maturities on the market  for a given name has increased, then one can say that, in the numerical examples described in this section, increased liquidity substantially reduces riskiness as measured by the maximum possible loss.

Consider a given starting portfolio $\alpha^{Old}$ of CDSs, and suppose  that there are a non-zero number of CDSs on the liquid market that one can use to hedge this starting portfolio.  Then the application of the procedure (\ref{optimization}) determines the hedge $\tilde{\alpha} + \beta$ and the hedged position $\Delta = \alpha^{Old} + \tilde{\alpha} + \beta$. The maximum possible loss $L_{Max}^H$ (where the superscript H indicates that this is for the hedged position) is then determined using (\ref{LMax}).  Now go through the same procedure, but take $\tilde{\alpha} = 0$, i.e. do not make use of the possibility to include CDSs from the liquid market in the hedge. One can nevertheless apply the optimization procedure of (\ref{optimization}) to obtain a $\beta$ such that the hedged position, which is now $\Delta = \alpha^{Old} + \beta$, satisfies the constraint of (\ref{optimization}).  The position $\Delta = \alpha^{Old} + \beta$ obtained in this case is called the `unhedged' position, where the quotes in `unhedged' mean that only $\beta$ (and no CDSs from the liquid market) has been used to obtain the hedged position. The value of $L_{Max}$ computed by this procedure is called $L_{Max}^U$, where the superscript $U$ stands for `unhedged'.

\begin{figure}
\includegraphics[scale = 0.7]{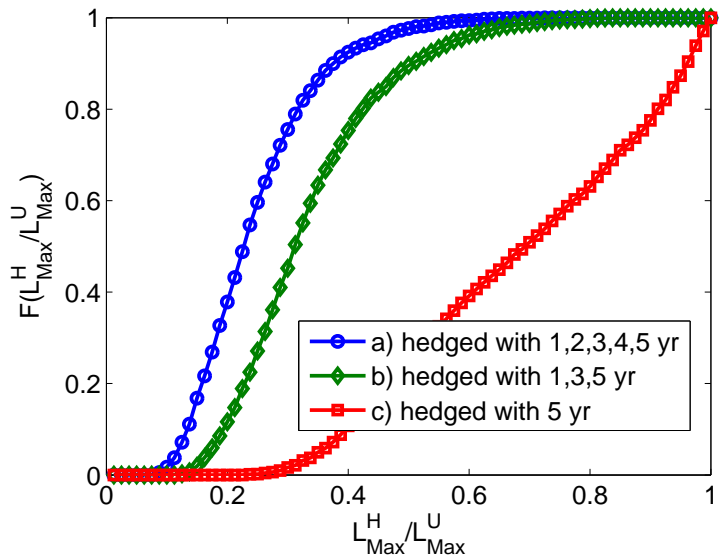} 
\caption{The cumulative distribution function $F(L_{Max}^H/L_{Max}^U)$ is plotted versus $L_{Max}^H/L_{Max}^U$ for three different assumptions for the maturities available on the liquid market, as indicated in the legend.}
\label{distributionFunction}
\end{figure}

The main objective of this subsection is to demonstrate in numerical examples, where there are a number of different CDS maturities on the liquid market that are available for hedging, that a hedged portfolio corresponding to a given starting portfolio $\alpha^{Old}$ can show a substantially lower value of the maximum possible loss, $L_{Max}$, than does the corresponding unhedged portfolio. This reduction in $L_{Max}$ by hedging depends on the starting  portfolio $\alpha^{Old}$.  In order to avoid the criticism that the starting portfolios studied were selected to be those that displayed the desired effect, the starting portfolios $\alpha^{Old}$ will be generated randomly.  Furthermore, results obtained for the cumulative distribution function $F(L^H_{Max}/L^U_{Max})$ defined below are more general, and are therefore of greater interest, than results obtained by studying a few selected starting portfolios.  

To implement the procedure just described, an initial  portfolio is defined by giving a value to the net notional $\alpha^{Old}_m$ corresponding to each termination time $T_m,\ \ m = 1,2,\dots,N$. The $N$ values of  $\alpha^{Old}_m$ are generated by drawing a sequence of $N$ pseudo-random numbers from the uniform distribution on the interval [-1,1].  Then $L_{Max}^H$ ($H$ for hedged) and $L_{Max}^U$ ($U$ for `unhedged') are calculated and the ratio $L_{Max}^H/L_{Max}^U$ is stored in the computer memory.  This process was repeated $nTrials$ times giving a set of $nTrials$ values of $L_{Max}^H/L_{Max}^U$ associated with the $nTrials$ randomly generated starting portfolios.  

A cumulative distribution function $F(L_{Max}^H/L_{Max}^U)$ for the random variable corresponding to  $L_{Max}^H/L_{Max}^U$ can be generated from the data stored in the computer memory (see previous paragraph).  The distribution function $F(L_{Max}^H/L_{Max}^U)$ is the probability that the random variable corresponding to $L_{Max}^H/L_{Max}^U$ has a value less than $L_{Max}^H/L_{Max}^U$.  This distribution function is shown in figure \ref{distributionFunction} for three different liquid markets of CDSs on the underlying name, liquid market a), which has maturities of 1, 2, 3, 4, and 5 years, liquid market b), which has maturities of 1, 3, and 5 years, and liquid market c) which has a single maturity of 5 years.  The same set of $nTrials$ randomly generated starting portfolios $\alpha^{Old}$ was used for the calculations on all three markets.  The upfront prices used for these maturities are those given in table \ref{data}, and the other parameters needed for calculation are given in section \ref{parameters}.

The first thing to note from Figure \ref{distributionFunction}, together with the data used to obtain it, is that none of the trials gives a value of $L_{Max}^H/L_{Max}^U$ that is greater than unity.  Also, for case a) for example, 100\% of the generated values for $L_{Max}^H/L_{Max}^U$ are less than 0.725, and 50\% of the generated values are less than 0.235.  The expected values of $L_{Max}^H/L_{Max}^U$ are, for case a) 0.244, case b) 0.333, and case c) 0.683.  There is thus quite a striking reduction in the expected maximum possible loss as a result of hedging.  In the case considered, the greater the number of maturities available on the liquid market, the greater this reduction.

\begin{figure}
\includegraphics[scale = 0.7]{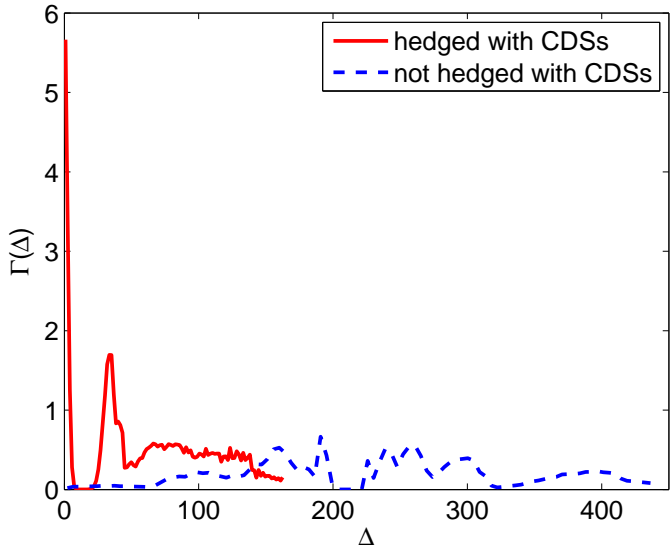}
\caption{For the `unhedged' portfolio defined by equation (\ref{alphaOld}), $L_{max}^U$ = 185.9\%, while for portfolio hedged with CDSs on the market having maturities 1, 2, 3, 4, and 5,  $L_{max}^H$ = 42.4\%.  The figure shows only the continuous part of the probability density $\Gamma(\Delta)$ for the two cases. There is also a discrete part of the probability density for which $\Delta = \Delta_0$ = 224.4\% with probability $S_0$ = 15.4\%  for the `unhedged' case, and $\Delta = \Delta_0$ = 0.0\% with probability $S_0 =$ 15.4\% for the hedged case.}
\label{GammaDelta}
\end{figure}

\begin{figure}  
\includegraphics[scale = 0.7]{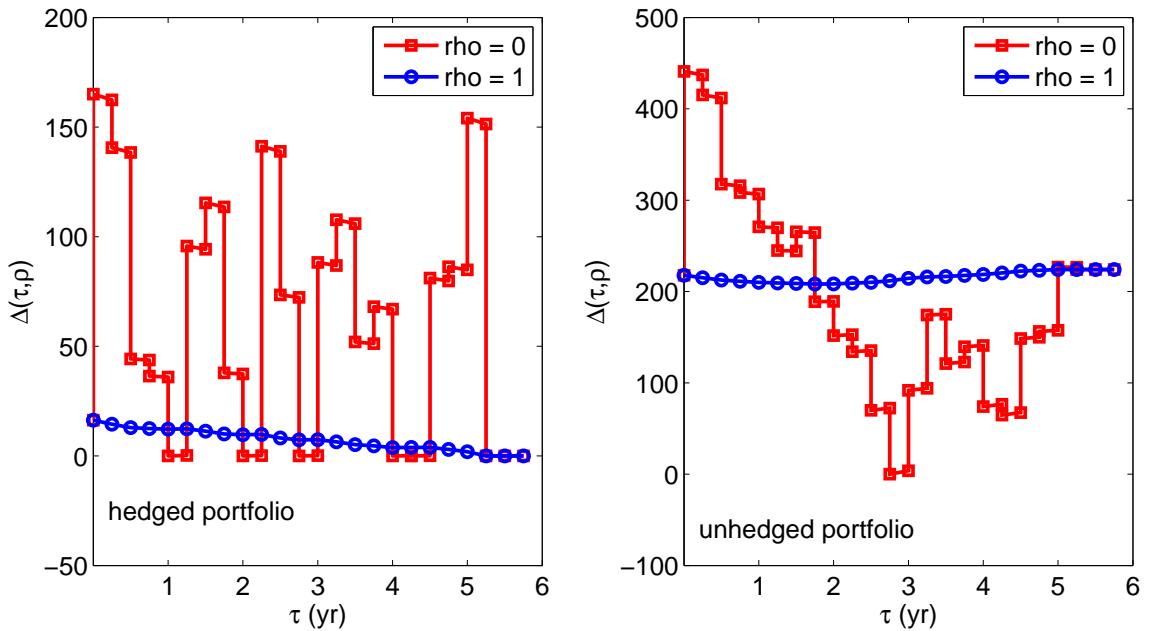}
\caption{Plot of $\Delta(\tau,\rho)$ for hedged and 'unhedged' (i.e. hedged only with $\beta$) portfolios.   The width in $\Delta$ for $\tau$ varying over a quarter (of a year) at fixed $\rho$ averages about 1.1 \% for the hedged case and 1.6 \% for the 'unhedged' case. }
\label{Delta_tau}
\end{figure}

\subsection{Probability Density for $\Delta$}\label{PDPL}
This subsection describes an example calculation of the probability density $\Gamma(\Delta)$ for $\Delta$, or, equivalently, the probability density for the profit and loss $\Psi$ (since $\Delta$ and $\Psi$ differ by an additive constant) and displays the result graphically.  The calculation is carried out for an example of one of the randomly generated starting portfolios $\alpha^{Old}$ that is given by (\ref{alphaOld}).  The continuous part of the probability density $\Gamma(\Delta)$ for $\Delta$, where $\Delta$ is the random variable given by (\ref{Delta}), is plotted in figure \ref{GammaDelta} for both the hedged and the  `unhedged' portfolios for case a) of figure \ref{distributionFunction}.  (See \citet{wal12} for details of a method of calculating $\Gamma(\Delta)$.)  There is also a discrete part of the probability density that is described in the caption. 

The plots for both the `unhedged' and hedged portfolios have considerable structure. Furthermore, they will both be quite asymmetric with respect to their mean values $\overline{\Delta}$, whatever these mean values are.  Thus, these probability densities have features that no simple description parameterized by only a few parameters can adequately capture.  Nevertheless, it is quite clear in figure \ref{GammaDelta} that the range of values of $\Delta$ that make a meaningful contribution to the the total probability is significantly greater for the `unhedged' portfolio than for the hedged portfolio. In other words, the `unhedged' portfolio is significantly riskier than the hedged portfolio.

The probability density for the profit and loss variable $\Psi(\tau,\rho)$ defined in (\ref{Psi}) is easily obtained from that for $\Delta(\tau,\rho)$ by translating the latter (shown in figure~\ref{GammaDelta}) in the direction of negative $\Delta$ by $\lambda\overline{\Delta}$.

A qualitative understanding of some features of the probability density $\Gamma(\Delta)$ can be obtained from a plot of the function $\Delta(\tau,\rho)$, which is shown in figure \ref{Delta_tau}.  From (\ref{Delta}), it follows that
\begin{equation}
	\Delta(\tau,\rho) = \Delta(\tau,\rho = 1) + (1-\rho)
		\left(\Delta(\tau,\rho = 0) - \Delta(\tau,\rho = 1)\right).
	\label{Delta2}
\end{equation}
The lines $\Delta(\tau,\rho = 0)$ and $\Delta(\tau,\rho = 1)$ are shown in the figure, and it follows from (\ref{Delta2}) that, at any given $\tau$, $\Delta(\tau,\rho)$ can be obtained by linear interpolation between $\Delta(\tau,\rho = 0)$ and $\Delta(\tau,\rho=1)$.  The function $\Delta(\tau,\rho)$ has discontinuities whenever $\tau$ passes from $\tau = T_m$ to $\tau = T_m + \epsilon$ where $T_m$ is the termination date of one of the CDSs in the portfolio, and $\epsilon$ is a positive infinitesimal. The reason for this is that the CDS with termination time $T_m$ makes a contribution to $\Delta(\tau,\rho)$ due to the loss on default.  This contribution is proportional to $(1-\rho)$ for $\tau \le T_m$, but is zero for $\tau > T_m$.

Note in figure \ref{GammaDelta} that the plot of $\Gamma(\Delta)$ versus $\Delta$ for the portfolio hedged with CDSs from the liquid market displays a high peak in $\Gamma(\Delta)$ for $\Delta$ close to $\Delta = 0$. Also note in figure \ref{Delta_tau} that for the hedged portfolio, for $\tau$ in the intervals $(1,1\frac{1}{4}]$, $(2,2\frac{1}{4}]$, $(2\frac{3}{4},3]$, $(4,4\frac{1}{4}]$ and $(4\frac{1}{4},4\frac{1}{2}]$ years, the lines $\Delta(\tau,\rho = 0)$ and $\Delta(\tau,\rho = 1)$ are very close together, and also very close to $\Delta = 0$.  This means that, for these ranges of $\tau$, all values of $\rho \in [0,1]$, weighted with their probability, contribute to approximately the same value of $\Delta \approx 0$, and the probability of having $\Delta$ in this small range of values close to zero is very high. On the other hand, for $\tau \in (0,\frac{1}{4}]$, the variation of $\rho$ from 1 to 0 causes $\Delta(\tau,\rho)$ to vary from about 15\% to 160\%.  This gives a much broader contribution of much lower height than that of the strong peak near $\Delta = 0$ described earlier, with most of its weight occurring in the region from $\Delta = 110\%$ to 160\% where the probability density $\gamma(\rho)$ has its greatest weight. It is therefore not surprising that $\Gamma(\Delta)$ exhibits considerable structure, when it is a superposition of a  number of disparate components such a those just discussed.  Also, for example, knowing the contribution of $\Delta(\tau,\rho)$ to $\Gamma(\Delta)$ for particular values of $\tau$ and $\rho$ will be of interest in understanding what happens in a scenario in which the default time is $\tau$ and the recovery rate is $\rho$.  The main point of this discussion is to note that, in assessing the risks associated with a given CDS position, one may find, in examining various scenarios, that the most serious risks occur for $(\tau,\rho)$ in a certain localized region.  This region will then give losses in a corresponding range of $\Delta$, (or equivalently, a corresponding range of profit and loss, $\Psi$).

It is important to know something about the general features of the probability density of $\Delta(\tau,\rho)$, the random variable that gives the present value of the payoff streams of a single-name CDS portfolio as a function of the path $(\tau,\rho)$.  It will be by discovering how to capture the features of $\Gamma(\Delta)$ that are essential to the determination of price movements of CDSs, that one will begin to make progress in the risk management of such portfolios. 

\subsection{Probability Density for $\Delta$ for a Constant Recovery Rate}\label{rhoConstPDPL}
Although it is universally accepted that the recovery rate is best described as a random variable whose value is only revealed after default, many  papers in the literature on the subject of valuing CDSs use the market-standard approach, take the recovery rate to be a constant, and do not calculate $\Gamma(\Delta)$. It is therefore of interest to investigate what changes might occur in $\Gamma(\Delta)$ as calculated in the previous section, in the case that the recovery rate is taken to be a constant, say $\rho_c$.  As in the previous section, the numerical calculations are carried out for the portfolio given in (\ref{alphaOld}).

\begin{figure}
\includegraphics[scale = 0.7]{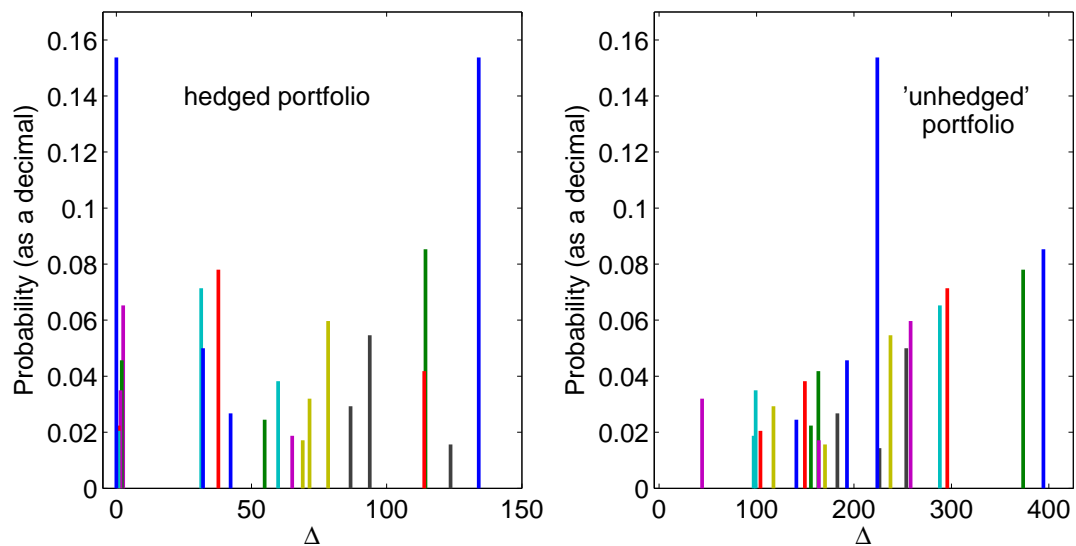}
\caption{When the recovery rate is taken to have a constant value, $\rho = \rho_c$, and when one considers a single-name porfolio of CDSs with contract maturities at the end of nearly every quarter, the probability density for $\Delta$ is well and simply approximated as that of a random variable with a discrete spectrum of values.}
\label{ProbVsDelta}
\end{figure}

\vspace{24pt}

When the recovery rate $\rho = \rho_c$, the value of $\Delta(\tau,\rho_c)$ can be found from (\ref{Delta2}). From this result, and from figure \ref{Delta_tau}, it is clear that, in any given quarter, say the $k$th where $\tau \in (T_{k-1},T_k]$, $\Delta(\tau,\rho_c)$ is approximately constant, and can be well approximated by 
\begin{equation}
	\Delta_k = 	=\frac{1}{2}(\Delta(T_{k-1}
	+\epsilon,\rho_c)+\Delta(T_k,\rho_c)),  
	\label{Dk}
\end{equation}		
where $\epsilon$ is a positive infinitesimal.	When the probability density for $\tau$ is described in terms of a constant hazard rate $h$, as in section \ref{parameters}, the probability of a default occurring in quarter $k$ is given by
\begin{equation}
	P_k = \exp{(-hT_{k-1})} -\exp{(-hT_k)}.
\label{Pk}	
\end{equation}
Thus the probability density for $\Delta$ can be accurately approximated by the the discrete spectrum shown in
figure \ref{ProbVsDelta}, where the probability of default in quarter $k$, $P_k$, is plotted versus the value of $\Delta$ in that quarter, $\Delta_k$.  The effect of allowing the recovery rate to be a random variable, is to smooth out the discrete spectrum, which then becomes the more or less smooth spectrum of figure \ref{GammaDelta}.

\section{Summary and Conclusions}\label{summary}
\begin{enumerate}
\item  An incomplete-market model designed to treat illiquid CDSs was introduced in \citet{wal12}. Because illiquid CDSs are not marketable (i.e. nonreplicable), risk-neutral pricing theory shows that these derivatives do not have a unique value, but have a range of fair prices located on one side of a no-arbitrage bound. The bound has a preference-free value.  Risk-neutral pricing theory does not determine a definite value for the price within the arbitrage-free range.  This must be determined by both the buyer and the seller balancing the risk versus return of the illiquid derivative while also taking account of their investment objectives and their ability to negotiate a successful outcome.  Thus, valuing an illiquid CDS is a combined modelling and risk-management task  (see, for example \citet{ros11} and \citet{ned11}): this article discusses almost exclusively only the first part of the exercise.

\item  Incomplete market techniques were used to assess the soundness of the approach followed in the market-standard model \citep{oka03}.  Illiquid CDSs were shown not to be marketable. This lack of marketability means that the risk-neutral valuation principle, which is used in the market-standard valuation process, is not applicable. Also, the correct application of incomplete-market ideas shows that, whereas the market-standard model finds a unique price for an illiquid CDS, the correct result is that there is an arbitrage free range of possible prices.  A combined modelling and risk management approach is therefore recommended for the valuation process.
 
\item  An elementary hedge called the vanilla hedge, is shown to reproduce analytically the principle features of the plot of no-arbitrage bounds versus illiquid CDS maturity. It accurately describes the features of this plot in an elementary, convincing and intuitive fashion, and is ideal for developing an understanding of no-arbitrage bounds.  For a given maturity, the results for the bounds are compared with a price for each illiquid maturity obtained by linear interpolation between the market prices.  The bounds on bid and ask prices can be very asymmetrically placed with respect to the line of interpolation so that the average of the bid and ask bounds for a given illiquid maturity is not necessarily close to the line of interpolation.

\item The incomplete-market approach, which was originally developed to study single CDSs, is extended to the study of single-name CDS portfolios. Studies of the effect of hedging a single-name CDS portfolio consisting of CDSs of maturity $T_m$ having notional $\alpha_m$, $m = 1,2,\cdots,N$ are carried out. These are hedged with the available liquid CDSs.  The notionals of the component CDSs of the portfolio are generated randomly and the results are characterized by determining the maximum possible loss, which is used as a measure of risk.  The randomness of the notionals is meant to mimic the randomness that might exist in the holding of a trading desk for a given name.  The hedging is remarkably effective in reducing the risk of holding the hedged position.  Also the greater the number of CDS maturities on the liquid market, the greater the risk reduction.

\item  The probability density $\Gamma(\Delta)$, where $\Delta$ is the random variable corresponding to $\Delta(\tau,\rho)$, the realized present value for the payoff stream corresponding to the path $(\tau,\rho)$, is plotted as a function of $\Delta$ for one of the single-name CDS portfolios that was randomly generated for the previous example.  There is significant structure in this function, which is a lot less spread out for the hedged position than for the unhedged position, indicating in a qualitative way that hedging reduces the risk.

\item  The probability density $\Gamma(\Delta)$ is found for the case where the recovery rate is taken to be constant (an assumption that is often made by practioners) and is shown to be well-described by a discrete spectrum.  This is qualitatively different from the case where the probability density for the recovery rate $\rho$ is taken to be a smooth function of $\rho$.

\item An evolution towards the general use of soundly based incomplete-market methods is desirable.  Only the simplest static hedge has been studied in this article and the previous one, and there remains much to be done. 
\end{enumerate}

\begin{appendix}

\section{Numerical Values of Parameters}\label{parameters}
Table \ref{data} lists the numerical values of certain parameters used in the numerical calculations described in this article.
\begin{table}
\begin{tabular}{|l|l|}
\hline
Quantity & Value \\ \hline 
running spread of all CDSs & $w = 500$ bp/yr \\
Nominal maturities of market CDSs and & 1, 2, 3, 4, and 5 yr \\
\ \ \ corresponding upfront prices & 5.25, 12.47, 18.08, 21.56, and 24.05 \% \\
risk-free interest rate & $r_F = 2 \%$ \\
physical probability of default within 1 year & $PD_1 = 30\%$ \\
expected return on capital at risk & $R_T = 25 \%$  \\ 
CDS contracts are concluded at time $t = T_0 = 0$;& \\
The difference between $T_1$ (the first premium payment time)& \\
\hspace{1 cm}	and $T_0$ is one quarter year&  \\ 
\hline
\end{tabular}
\caption{Values of parameters used in numerical examples (unless otherwise stated).}
\label{data} 
\end{table} 

Also, in order to find the statistical properties of the portfolios of interest, it is necessary to specify the physical probability density for the default time $\tau$ (called $\Upsilon(\tau)$), and the physical probability density for the recovery rate $\rho$ given a particular default time $\tau$, called $\gamma(\rho|\tau)$. A simplifying assumption made in this article is that $\tau$ and $\rho$ are independent random variables,  so that $\gamma(\rho|\tau) = \gamma(\rho)$.  The probability density $\gamma(\rho)$ is taken to be that segment of a normal probability density of mean $\mu = 0.15$ and standard deviation $\sigma = 0.16$ existing for $\rho \in [0,1]$.
The probability density for the default time, $\Upsilon(\tau)$, is specified in terms of a constant hazard rate $h$ so that $\Upsilon(\tau) = h\exp{(-h\tau)},\ \ 0 \leq \tau \leq T_N,\ $. Also,the hazard rate $h$ is simply estimated in terms of the probability of default within one year, $PD_1$, using $h = -\log(1-PD_1)$.  There is no difficulty in principle to specifying $\Upsilon(\tau)$ and $\gamma(\rho|\tau)$ numerically to have any desired form.

An example randomly generated portfolio for which the probility density $\Gamma(\Delta)$ of the realized present value $\Delta$ of the payoff streams was calculated above.
\begin{eqnarray}
	\alpha^{Old} &=& [0.2190\ \ 0.9513\ \ 0.0744\ \ 0.3669 \ \ 0.2543      \ \ -0.2179\ \ 0.7840\ \ 0.3894\nonumber \\ 
	& &0.1945\ \ 0.6885\ \ 0.7642\ \ -0.9360\ \ -0.8572\ \ 0.5786\ \ -0.1819 \nonumber \\ 
	& &0.7254\ \ 0.1285\ \ -0.8874\ \ -0.0712\ \ -0.7650\ \ 0.0290].
	\label{alphaOld}
\end{eqnarray}

\end{appendix}
\end{document}